\documentclass{ws-ijmpd}

\usepackage[super,compress]{cite}
\usepackage[breaklinks]{hyperref}
\hypersetup{colorlinks,urlcolor=black,citecolor=black,linkcolor=black,filecolor=black}
\usepackage{amsmath,amssymb}
\usepackage{latexsym}
\usepackage{slashed}
\usepackage{arydshln}
\usepackage{braket}
\allowdisplaybreaks

\usepackage{graphicx}

\usepackage[sharp]{easylist}
\newcommand{\fmetric}{\mathcal{G}}

\newcommand{\abs}[1]{\left| #1 \right|}

\graphicspath{{../images/}}

\usepackage{amssymb}
\begin{document}

% \begin{frontmatter}
\title{Exploring the inflation of $F(R)$ gravity}

% \author[1,2,3]{TOMOHIRO INAGAKI}
\author{TOMOHIRO INAGAKI}
% \ead{inagaki@hiroshima-u.ac.jp}
% \address[1]{Information Media Center, Hiroshima University, Higashi-Hiroshima, 739-8521, Japan}
% \address[2]{Core of Research for the Energetic Universe, Hiroshima University, Higashi-Hiroshima, 739-8526, Japan} \address[3]{Lab. Theor. Cosmology, Tomsk State University of Control Systems and Radioelectronics, 634050 Tomsk, Russia}
\address{Information Media Center, Hiroshima University, Higashi-Hiroshima, 739-8521, Japan}
\address{Core of Research for the Energetic Universe, Hiroshima University, Higashi-Hiroshima, 739-8526, Japan}
\address{Lab. Theor. Cosmology, Tomsk State University of Control Systems and Radioelectronics, 634050 Tomsk, Russia}
%
%
% \author[4]{HIROKI SAKAMOTO}
\author{HIROKI SAKAMOTO}
% \ead{h-sakamoto@hiroshima-u.ac.jp}
% \address[4]{Department of Physics, Hiroshima University, Higashi-Hiroshima, 739-8526, Japan}
\address{Department of Physics, Hiroshima University, Higashi-Hiroshima, 739-8526, Japan}

% \author{FIRST AUTHOR\footnote{Typeset names in
% 8~pt roman, uppercase. Use the footnote to indicate the
% present or permanent address of the author.}}
%
% \address{University Department, University Name, Address\\
% City, State ZIP/Zone,
% Country\footnote{State completely without abbreviations, the
% affiliation and mailing address, including country. Typeset in
% 8~pt Times italic.}\\
% first\_author@university.edu}

\maketitle

\begin{abstract}
The early time expansion of the space-time, namely inflation, is introduced to solve some cosmological problems.
$F(R)$ gravity is a simple extension of the general relativity to induce the inflationary expansion.
The precise observation of the Cosmic Microwave Background gives us the information to inspect the model of the inflation.
By the Weyl transformation, $F(R)$ gravity can be transformed to the Einstein-Hilbert term with a scalar field that plays the role of the inflaton.
In this paper we obtain general formulae to derive the inflationary parameters including the models with a non-canonical kinetic term.
The inflationary parameters are described as functions of the inflation potential and its derivatives.
We apply the procedure to $F(R)=R+\gamma R^n$ and $R(1+\gamma R^n)^{1/n}$ models and show the allowed parameters space.
\end{abstract}

% \begin{keyword}
% 	Modified gravity, Inflation, CMB fluctuations
% \end{keyword}

% \end{frontmatter}

% ---------------------------------------------
% 1. introduction
% ---------------------------------------------
\section{Introduction}
The discovery of the gravitational wave from the heavy celestial body and
the direct observation of the Black Hole are great success for the Einstein's general relativity.
From the cosmological point of view, however, there are some remaining problems, e.g.~the flatness, horizon, monopole problems.
The inflation theory %which is the accelerating expansion of the universe at the early-time epoch 
is one of elegant and simple solutions for these problems (for reviews, see, e.g. Refs.~\refcite{Langlois:2010xc,Linde:2014nna,Baumann:2014nda}).
Introducing an inflaton field
with a plateau like potential, the non-vanishing potential energy can induce the accelerating expansion of the space-time and provide an enough e-folding number.
The alternative approach is found in the modification of the Einstein gravity.
A simple and useful model is $F(R)$ gravity where the Einstein-Hilbert action is replaced by a general function of Ricci scalar $R$.
The model can introduce the accelerating expansion not only the inflation era but also the present universe,
see Refs.~\refcite{Clifton:2005aj,Nojiri:2006ri,Nojiri:2010wj,Clifton:2011jh,Nojiri:2017ncd,Capozziello:2019cav}.
It is known that the accelerating expansion takes place by the $R^2$ term for a large curvature~\cite{Starobinsky:1980te} and the resulting inflationary parameters satisfy the observable constraints for the Cosmic Microwave Background (CMB)~\cite{Akrami:2018odb,Akrami:2019izv}.

For $F(R)$ gravity, an extra degree of freedom plays a role of the inflaton and there are a variety of models.
Recent observation of the CMB and its fluctuations increases accuracy enough to inspect the models.
Thus we launch a plan to develop a simpler procedure is required to evaluate the CMB fluctuations for general models of $F(R)$ gravity.
The $F(R)$ term is simplified to the Einstein-Hilbert term with a scalar field by the Weyl transformation.
The complexity to calculate the inflationary parameters comes from the definition of the scalar field with a canonical kinetic term.
For a non-canonical kinetic term it is possible to reduce the expression for the inflaton potential.
The coefficient of the non-canonical kinetic term can be written as the metric of configuration space,
see Refs.~\refcite{Sasaki:1995aw,Peterson:2011yt,Gao:2014fva,Elliston:2012ab}.
Here, we calculate the inflationary parameters for the models
with a non-canonical kinetic term.
It is noted that another technique is developed to analyze $F(R)$ gravity without using the Weyl transformation~\cite{Brooker:2016oqa}.

The paper is organized as follows: in \autoref{sec:covariant}, general formulae for the inflationary parameters are derived for $F(R)$ gravity by using the covariant approach. In \autoref{sec:application} the formulae are used to calculate the inflationary parameters in $F(R)=R+\gamma R^{n}$ and $R{(1+\gamma R^{n})}^{1/n}$ models and show the constraint for the model parameters. The \autoref{sec:concludion} is devoted for the concluding remarks.

% --------------------------------------------------------------------
% section 2
% --------------------------------------------------------------------
\section{Covariant approach in $F(R)$ gravity}\label{sec:covariant}

$F(R)$ gravity is defined by the action,
% The action is
\begin{align}
	S = \int d^4x \sqrt{-g} \Big[ \frac{F(R)}{2\kappa^2} \Big],
	\label{act:jordan}
\end{align}
where $\kappa$ denotes the inverse of the Planck mass, $F(R)$ is an analytic function of the Ricci scalar $R$.
For practical calculations of the inflationary parameters, it is more convenient to introduce an auxiliary field, $A$, called scalaron
and consider the following equivalent action,
\begin{align}
	S = \int d^4x \sqrt{-g} \Big[ \frac{F_A \cdot (R - A) + F(A)}{2\kappa^2} \Big],
	\label{act:aux}
\end{align}
where we use the notation that the lower index $A$ denotes the ordinary derivative.
Substituting the equation of motion for $A$ into the action (\ref{act:aux}), 
we reproduce the original action (\ref{act:jordan}).

To simplify the calculations we move to the Einstein frame by the Weyl transformation, $g_{\mu\nu} \to F_{A}^{-1} g_{\mu\nu}$, and obtain
\begin{align}
	S &= \int d^4x \sqrt{-g} \Big[ \frac{R}{2\kappa^2}
		- \frac{3}{4\kappa^2} (F_A^{-1} \partial F_A)^2
	- \frac{AF_A - F(A)}{2\kappa^2 F_A^2} \Big],
	\label{act:einstein}
\end{align}
where the determinant of the metric tensor $g$ and the Ricci scalar $R$ represent
the ones in the transformed space-time %. In \eqref{act:einstein} 
and the total derivative term is dropped.
The kinetic term for the scalaron $A$ is written as
\begin{align}
	- \frac{3}{4\kappa^2} (F_A^{-1} \partial F_A)^2
	= - \frac{1}{2} \fmetric_{AA} (\partial_\mu A) (\partial^\mu A),
\end{align}
with the field-space metric,
\begin{align}
	\fmetric_{AA} &= \frac{3}{2\kappa^2} \Bigl( \frac{F_{AA}}{F_A} \Bigr)^2.
	\label{metric:field_space}
\end{align}
The potential for the scalaron $A$ is defined by
\begin{align}
	V \equiv \frac{AF_A - F}{2\kappa^2 F_A^2}.
	\label{eq:potential_fr}
\end{align}
Though the scalaron $A$ can be redefined to acquire the canonical kinetic term, an explicit expression for the potential~\eqref{eq:potential_fr} is not always obtained as a function of the redefined field. 
To evaluate the model with a non-canonical kinetic term,
we employ the covariant approach~\cite{Sasaki:1995aw,Peterson:2011yt,Gao:2014fva,Elliston:2012ab}.

We consider the following action with $\mathcal{N}$ species of scalar fields,
\begin{align}
	S = \int d^{4}x \sqrt{-g} \Bigl[ \frac{R}{2\kappa^2} - \frac{1}{2} \fmetric_{IJ} \partial_\mu \phi^I \partial^\mu \phi^J - V(\phi^I) \Bigr],
	\label{eq:action-gen-kinetic-KG}
\end{align}
where %Ricci scalar $R$, $\kappa = 1/M_{Pl}$, 
the field-space metric $\fmetric_{IJ}$ is a symmetric function of the scalar fields, 
$\phi^I (I, J = 1, 2, \dots, \mathcal{N})$.% and $V(\phi^I)$ is the potential term.

In the large class of inflation models it is assumed that the energy density of the universe is dominated by the potential energy, $V$, of the slowly-rolling scalar fields. To derive the slow-roll parameters, we first introduce the model-independent parameters,
called the horizon flow functions~\cite{Schwarz:2001vv}, along to the geodesic direction,
\begin{align}
	\epsilon_{n+1} = \frac{1}{H} \mathcal{D}_t \ln \abs{\epsilon_{n}},
\end{align}
where $H$ represents the Hubble parameter and we set $n \geq 0$ and $\epsilon_{0} \propto 1/H$.
The directional derivative $\mathcal{D}_{t} = \frac{d\phi^{I}}{dt} \nabla_{I}$ is defined with the covariant derivative $\nabla_{I}$ on the field-space manifold.

The slow-roll condition is satisfied for $\epsilon_{1} < 1$ and the other functions $\epsilon_{n} (n>1)$ are typically the same order of magnitude with $\epsilon_{1}$. Under the slow-roll condition Friedmann equation and the equation of motion for the background field $\phi^{I}(t)$ are given by
\begin{align}
	3H^2 = \kappa^2 V, \quad 3H\mathcal{D}_{t} \phi^I + V^I = 0.
	\label{eq:slow_roll_condition}
\end{align}
From \autoref{eq:slow_roll_condition}, the horizon flow functions are described by the slow-roll parameters, $\varepsilon_{V}, \eta_{V}$ and $\xi_{V}$,
\begin{align}
	\epsilon_{1} &= -\frac{\mathcal{D}_{t} H}{H^{2}} \sim \varepsilon_{V}, \\
	\epsilon_{2} &= \frac{\mathcal{D}_{t}^{2} H}{H \mathcal{D}_{t} H} - 2\frac{\mathcal{D}_{t} H}{H^{2}} \sim 4\varepsilon_{V} - 2 \eta_{V}, \\
	\epsilon_{2} \epsilon_{3} &\sim 
	2\xi_{V} - 16 \varepsilon_{V}^{2} + 6 \varepsilon_{V} \eta_{V},
\end{align}
with
\begin{align}
	&\varepsilon_V = \frac{1}{2\kappa^2} \frac{V_{I} V^{I}}{V^2},
	\label{eq:potential_slow_roll_1st} \\
	&\eta_V = \frac{V^{I} V^{J} V_{;IJ}}{\kappa^2 V V^{K} V_{K}},
	\label{eq:potential_slow_roll_2nd} \\
	&\xi_V = \frac{V^{I} V^{J} V^{K} V_{;IJK}}{\kappa^4 V^2 V^{M} V_{M}},
	\label{eq:potential_slow_roll_3rd}
\end{align}
where the semicolon indicates the covariant derivative on the field-space 
manifold: $V_{;IJ\cdots} \equiv \nabla_I \nabla_J \cdots V$,
and we write the first derivative as $V_{I} = V_{;I}$.% = \partial V / \partial \phi^I$.
The lower index are raised by the field-space metric, $V^{I} = \fmetric^{IJ} V_{J}$.
Using \autoref{eq:potential_slow_roll_1st}, the e-folding number, $N$, is descrived as
\begin{align}
	N \equiv \int H dt = \int_{C} d\phi^I \frac{ V_{I} }{ 2V\varepsilon_{V} },
	\label{eq:efold_cov}
\end{align}
where $C$ is a path from the horizon crossing to the end of inflation.

In the covariant approach, the two- and three-point correlation functions are also computed among the geodesic direction. In this paper we refer the results of the power spectrum, $A_s$, the spectrum index, $n_s$, the running of the spectrum index, $\alpha_{s}$, the tensor-to-scalar ratio, $r$, and the non-gaussianity, $f_\mathrm{NL}$, 
for details, see Refs.~\refcite{Sasaki:1995aw,Gong:2011uw,Elliston:2012ab,Gong:2014kpa}.
\begin{align}
	&A_s = \frac{\kappa^2 V}{12\pi^2} N^{I} N_{I},
	\label{obs:power_spectrum_multi_field} \\
	&n_{s} = 1 - 6\varepsilon_V + 2\eta_V,
	\label{obs:spectrum_index_multi_field} \\
	&\alpha_s = -24\varepsilon_V^2 + 16\varepsilon_V \eta_V - 2\xi_V,
	\label{obs:run_spectrum_index_multi_field} \\
	&r = 16\varepsilon_V,
	\label{obs:tensor_scalar_multi_field} \\
	&f_\mathrm{NL} = -\frac{5}{6} \frac{N^{I} N^{J} N_{;IJ}}{(N^K N_K)^2}.
	\label{obs:non_gaussianity_multi_field}
\end{align}
It should be noted that there are terms proportioanl to the Riemann tensor in the original papers. 
These terms disappears from Eqs.~\eqref{eq:slow_roll_condition} and \eqref{eq:efold_cov}. 

Here we assume that the potential of the scalaron in the $F(R)$ gravity dominates the energy density of the early universe and regard the scalaron  as the slow-roll scalar field. Substituting the field metric~\eqref{metric:field_space} and the potential~\eqref{eq:potential_fr} into~\eqref{eq:potential_slow_roll_1st}--\eqref{eq:potential_slow_roll_3rd}, we obtain the slow-roll parameters,
\begin{align}
	&\varepsilon_V = \frac{1}{3} \Bigl( \frac{2F - AF_{A}}{AF_{A} - F} \Bigr)^2, %\\
	\label{eq:slow-roll-1st-fr}\\
	&\eta_V = -\frac{2}{3} \frac{2F-AF_{A}}{AF_A-F} \Bigl( 2 - \frac{F_A}{F_{AA}} \frac{F_A-AF_{AA}}{2F-AF_A} \Bigr),
	\label{eq:slow-roll-2nd-fr}\\
	&\xi_V = \frac{4}{9} \Bigl( \frac{2F-AF_A}{AF_A-F} \Bigr)^2 \left[ 1 + \frac{6F - \frac{F_A^3 F_{AAA}}{F_{AA}^3} - 3\frac{F_A^2}{F_{AA}}}{2F-AF_A} \right],
	\label{eq:slow-roll-3rd-fr} %\\
\end{align}
and the e-folding number,
\begin{align}
	&N = \int^{A_N}_{A_e} dA \frac{3F_{AA}}{2F_{A}} \frac{AF_{A} - F}{2F - AF_{A}} .
	\label{eq:e-folds-fr}
\end{align}
The interval of integration for the e-folding number is given by the configuration at the horizon crossing, $A_N$, and  the one at the end of inflation, $A_e$.
The value, $A_e$, is fixed by the slow-roll condition, $\varepsilon_V = 1$. The value, $A_N$, is tuned to produce a suitable e-folding number.

Inserting Eqs.~\eqref{eq:slow-roll-1st-fr}--\eqref{eq:e-folds-fr} into Eqs.~\eqref{obs:power_spectrum_multi_field}--\eqref{obs:non_gaussianity_multi_field},
we find the explicit expression of the inflationary parameters,
\begin{align}
	&A_s = \frac{\kappa^{2} (A F_A  - F )^{3}}{16 \pi^{2} F_A^2 (A F_A - 2 F)^2},
	\label{obs:power_spectrum_fr} \\
	&r = \frac{16}{3} \Bigl( \frac{2F - AF_{A}}{AF_{A} - F} \Bigr)^2,
	\label{obs:tensor_scalar_fr}\\
	&n_{s} = 1 -\frac{2}{3} \frac{4F^2 - A^2F_{A}^2}{(AF_{A} - F)^2} + \frac{4}{3} \frac{F_{A}}{F_{AA}} \frac{F_{A}-AF_{AA}}{AF_{A}-F},
	\label{obs:spectrum_index_fr}\\
	&\begin{aligned}[b]
		\alpha_s =&
	-\frac{8}{3} \Bigl( \frac{2F - AF_{A}}{AF_{A} - F} \Bigr)^4
	% \\
				  -\frac{32}{9} \Bigl( 2 - \frac{F_A}{F_{AA}} \frac{F_A-AF_{AA}}{2F-AF_A} \Bigr) \Bigl( \frac{2F - AF_{A}}{AF_{A} - F} \Bigr)^3 \\
				  &-\frac{8}{9} \Bigl( \frac{2F-AF_A}{AF_A-F} \Bigr)^2 \Biggl[ 1 + \frac{6F - \frac{F_A^3 F_{AAA}}{F_{AA}^3} - 3\frac{F_A^2}{F_{AA}}}{2F-AF_A} \Biggr],
	\end{aligned}
	\label{obs:run_spectrum_index_fr}\\
	&\begin{aligned}[b]
		f_\mathrm{NL} =& \frac{10}{9} \Bigl( \frac{2F - AF_{A}}{AF_{A} - F} \Bigr)^2
		% \\
					   - \frac{5}{9}\frac{2F-AF_{A}}{AF_A-F} \Bigl( 2 - \frac{F_A}{F_{AA}} \frac{F_A-AF_{AA}}{2F-AF_A} \Bigr).
	\end{aligned}
	\label{obs:non_gaussianity_fr}
\end{align}
Therefore the inflationary parameters  in the $F(R)$ gravity are calculated by \autoref{eq:e-folds-fr} and Eqs.~\eqref{obs:power_spectrum_fr}--\eqref{obs:non_gaussianity_fr} under the slow-roll scenario.

% ------------------------------------------------------------------
% Section 3
% ------------------------------------------------------------------
\section{Inflationary parameters in \texorpdfstring{$F(R)$}{} gravity}
\label{sec:application}

In this section we calculate the inflationary parameters in two types of $F(R)$ gravity.
We first consider the power-law model to inspect the validity of our results \autoref{eq:e-folds-fr} and Eqs.~\eqref{obs:power_spectrum_fr}--\eqref{obs:non_gaussianity_fr}.
The inflation for power-law model was also studied in Ref.~\refcite{Sebastiani:2013eqa}.
Next we investigate a modified power-law model which has a no analytic canonical kinetic term.

\subsection{Power-law model}
\label{sec:power-law}
The power-law model is defined by
\begin{align}
	F(R) = R + \gamma R^n,
\end{align}
where $\gamma$ is a positive coupling with mass dimension, $2-2n$, and $n$ is a real number, not 1.
For $n\sim 2$ the model is considered as a quantum corrected $R^2$ model, see Refs.~\refcite{Motohashi:2014tra,Liu:2018htf,Ben-Dayan:2014isa,Odintsov:2017hbk}. 
At the limit $n\rightarrow 2$, the model is equal to the Starobinsky model~\cite{Starobinsky:1980te}.

In the power-law model the integration in \autoref{eq:e-folds-fr} is analytically performed and found
\begin{align}
	N = \frac{3n}{4(2-n)} \ln \abs{ \frac{1+(2-n)\gamma A_N^{n-1}}{1+(2-n)\gamma A_e^{n-1}} }
  - \frac{3}{4} \ln \abs{\frac{1+n\gamma A_N^{n-1}}{1+n\gamma A_e^{n-1}}} .
  \label{eq:e-fold-fr-power}
\end{align}
By solving the condition $\varepsilon_V = 1$, we obtain the field variable at the end of inflation,
\begin{align}
	\gamma A_e^{n-1} = \frac{1}{\sqrt{3}(n-1)+n-2}.
\end{align}

\begin{figure}[htbp]
	\centering
	\includegraphics[width=0.7\linewidth]{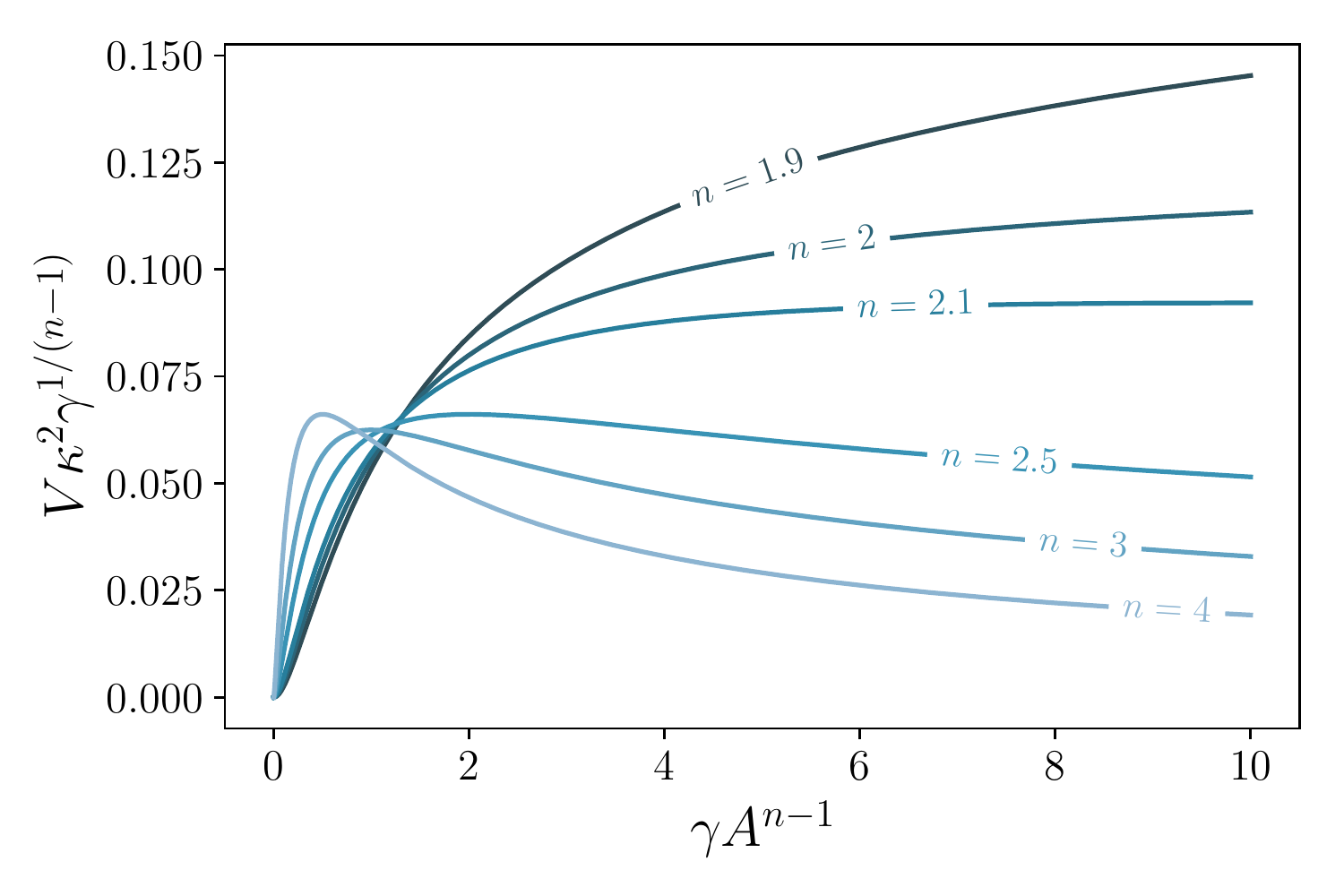}
	\caption{Behavior of the potential for the power-law model~\eqref{eq:power-law-potential} for $n=1.9, 2, 2.1, 2.5, 3, 4$.}
	\label{images:power_law_potential}
\end{figure}

From \autoref{eq:potential_fr} the potential of the power-law model is given by
\begin{align}
	V = \frac{A^{n + 2} \gamma \left(n - 1\right)}{2\kappa^2 \left(A + A^{n} \gamma n\right)^{2}}.
	\label{eq:power-law-potential}
\end{align}
In \autoref{images:power_law_potential} the typical behavior of the potential is shown as a function of the field variable $A$.
It is observed that the potential~\eqref{eq:power-law-potential} is bounded above for $n>2$. 
The slow-roll parameter, $\varepsilon_V$, vanishes and the integrand in \autoref{eq:e-folds-fr} diverges at the maximum of potential,
\begin{align}
	\gamma A_{\mathrm{max}}^{n-1} = \frac{1}{n-2}.
	\label{eq:field-config-max-fr-power}
\end{align}
To avoid the divergence we set the interval of the integration $A < A_{\mathrm{max}}$.
The field variable at the horizon crossing $A_N$ is calculated by solving \autoref{eq:e-fold-fr-power} with $N=60$.
The coefficient of the first term in \autoref{eq:e-fold-fr-power} becomes large at $n\sim 2$. 
Then it is found that the field variable at the horizon crossing, $A_N$, realized near $A_{\mathrm{max}}$ for $n>2$.
In these cases the main contribution to the e-folding number is comes from the first term in \autoref{eq:e-fold-fr-power}. 
Hence, we negrect the second term and $A_e$ in \autoref{eq:e-fold-fr-power} and find an approximate expression of $A_N$ which is valid for $n\gtrsim 2$,
\begin{align}
	\gamma A_N^{n-1} \sim \frac{1}{n-2} \Bigl( 1-e^{-\frac{4(n-2)N}{3n}} \Bigr).
	\label{eq:horizon_crossing_power_law}
\end{align}
Substituting \autoref{eq:horizon_crossing_power_law} into Eqs.~\eqref{obs:power_spectrum_fr}--\eqref{obs:non_gaussianity_fr}, we obtain the inflationary parameters as a function of the model parameter $n$,
\begin{align}
	&\begin{aligned}[b]
		A_s = \frac{\kappa^2}{16\pi^2 \nu_n^3}\Biggl( \frac{1}{\gamma (n-2)} \frac{2}{1+x} \Biggr)^{2/(n-1)}
		\Biggl( \frac{2}{1+x} \Biggr)^3 %\\
		  % \times 
		  \Biggl( 1+\frac{n}{n-2} \frac{2}{1+x} \Biggr)^{-1}
		  \Biggl( \frac{1+x}{1-x} \Biggr)^2,
	\end{aligned}
	\label{obs:power_spectrum_power_law} \\
	&n_s = 1 - \frac{2}{3} \nu_n^2 \Biggl[1 + \frac{1+x}{n-2} + \frac{3n-2}{4n}(1+x)^2\Biggr],
	\label{obs:spectrum_index_power_law} \\
	&r = \frac{4}{3} \nu_n^2 (1-x)^2,
	\label{obs:tensor_scalar_power_law} \\
	&\begin{aligned}[b]
		\alpha_s = \frac{8}{9} \nu_n^3 \Biggl[
				   &\frac{(-3n+2)(n-2)}{16n^2} (1+x)^4 % \\
				   - \frac{7n-4}{8n^2}(1+x)^3 \\
				   &+ \frac{3n^2-8n+2}{4n(n-2)}(1+x)^2 % \\
				   + \frac{1+x}{3(n-2)}
			   \Biggr],
	\end{aligned}
	\label{obs:run_spectrum_index_power_law} \\
	&\begin{aligned}[b]
		f_\mathrm{NL} = \frac{5}{9} \nu_n^2 \Biggl[
			&3 - \frac{(7n-13)(1+x)}{2(n-2)} %\\
					   + \frac{2n+1}{4n}(1+x)^2
				   \Biggr],
	\end{aligned}
	\label{obs:non_gaussianity_power_law}
\end{align}
with
\begin{align}
  \nu_n = (n-2)/(n-1), \quad x = \coth\frac{2(n-2)N}{3n} .
\end{align}

\begin{figure}[htbp]
	\centering
	\includegraphics[width=0.7\linewidth]{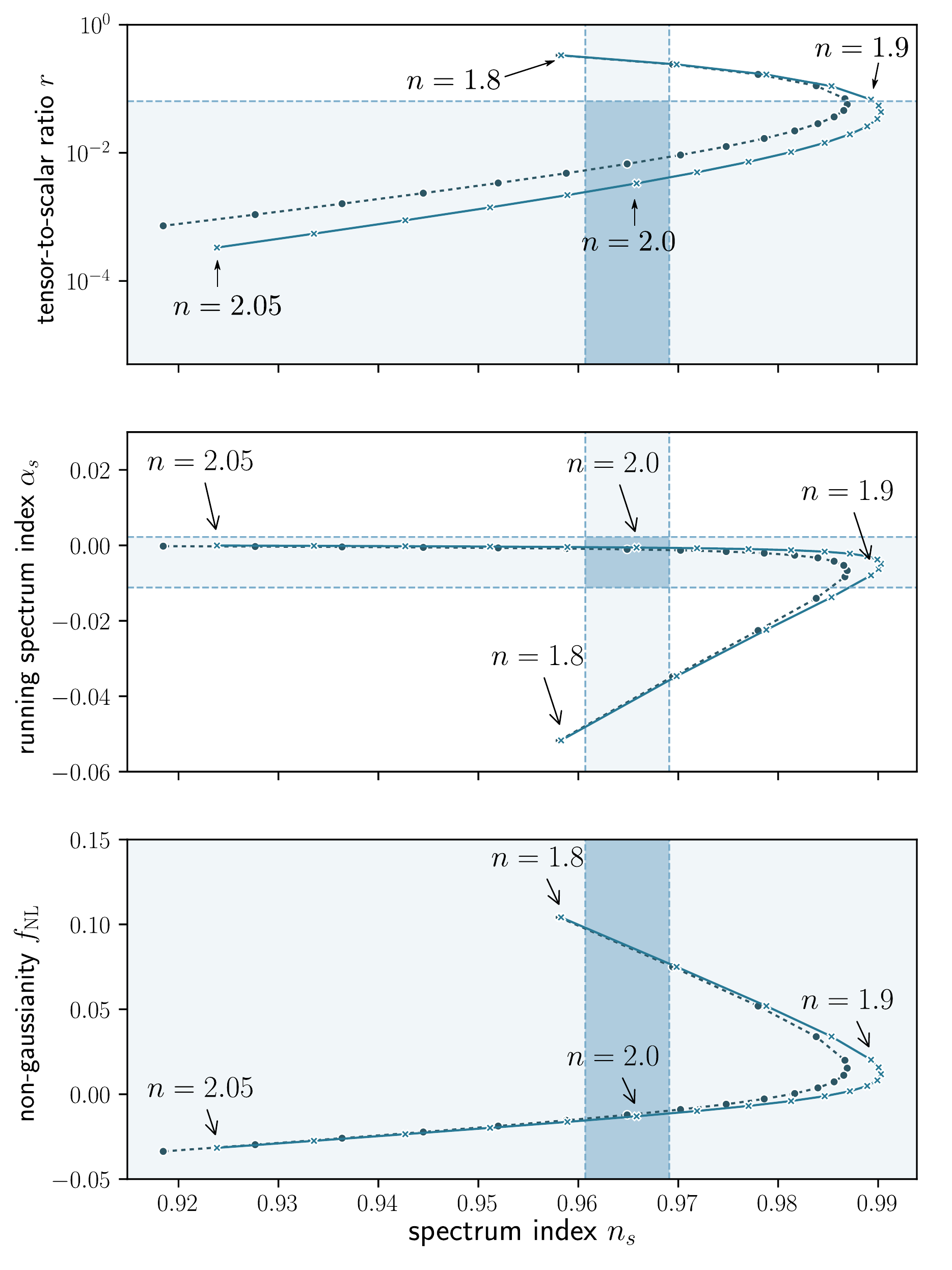}
	\caption{Inflationary parameters in the power-law model from $n=1.8$ to $n=2.05$ for $N=50$ (dashed dark blue lines) and $N=60$ (solid blue lines). The darker shade area shows the parameter range consistent with the Planck 2018 observations~\cite{Akrami:2018odb,Akrami:2019izv}.}
	\label{images:power-law-obs}
\end{figure}
In \autoref{images:power-law-obs} the inflationary parameters are shown between $n=1.8$ and $n=2.05$.
We observe that the power-law model is consistent with the Planck 2018 observations for $n \sim 2$ with $N=60$. 
At the limit $n\rightarrow 2$, the e-folding number reduces to be
$N \sim \frac{3}{2} \gamma A_N$. From Eqs.~\eqref{obs:power_spectrum_power_law}--\eqref{obs:non_gaussianity_power_law} the inflationary parameters 
are given by
\begin{align}
	\begin{aligned}
		&A_s = \frac{\kappa^2 N^4}{9\pi^2 \gamma (4N+3)^2}, \quad
		n_s = 1 - \frac{2}{N} - \frac{3}{N^2}, \quad
		r = \frac{12}{N^2}, \\
		&\alpha_s = -\frac{2}{N^2} - \frac{15}{2N^3} - \frac{9}{2N^4}, \quad
		f_\mathrm{NL} = -\frac{5}{6N} + \frac{25}{8N^2}. %\\
	\end{aligned}
	\label{eq:starobinsky-obs}
\end{align}
These results reproduce the ones in the Starobinsky inflation~\cite{Starobinsky:1980te}.

\subsection{Modified power-law model}
\label{sec:toy}
Next we modify the power-law model in the following form
\begin{align}
	F(R) = R(1 + \gamma R^n)^{1/n}.
	\label{eq:model-toy}
\end{align}
In the model
the kinetic term for the auxiliary field, $A$, has a non-canonical form.
Though it is always possible to transform the field variable and define the theory with a canonical kinetic term,
the transformed field can not be described as an analytic function of the original field for $n\neq 1$.
Thus the covariant approach developed in the previous section is absolutely necessary. 
It should be noted that the model is equivalent to the Starobinsky model at $n=1$.

The potential for the auxiliary field is given by
\begin{align}
	V = \frac{A^{n + 1} \gamma \left(A^{n} \gamma + 1\right)^{1 - 1/n}}{2 \kappa^{2} \left(2 A^{n} \gamma + 1\right)^2}.
\end{align}
\begin{figure}[htbp]
	\centering
	\includegraphics[width=0.7\linewidth]{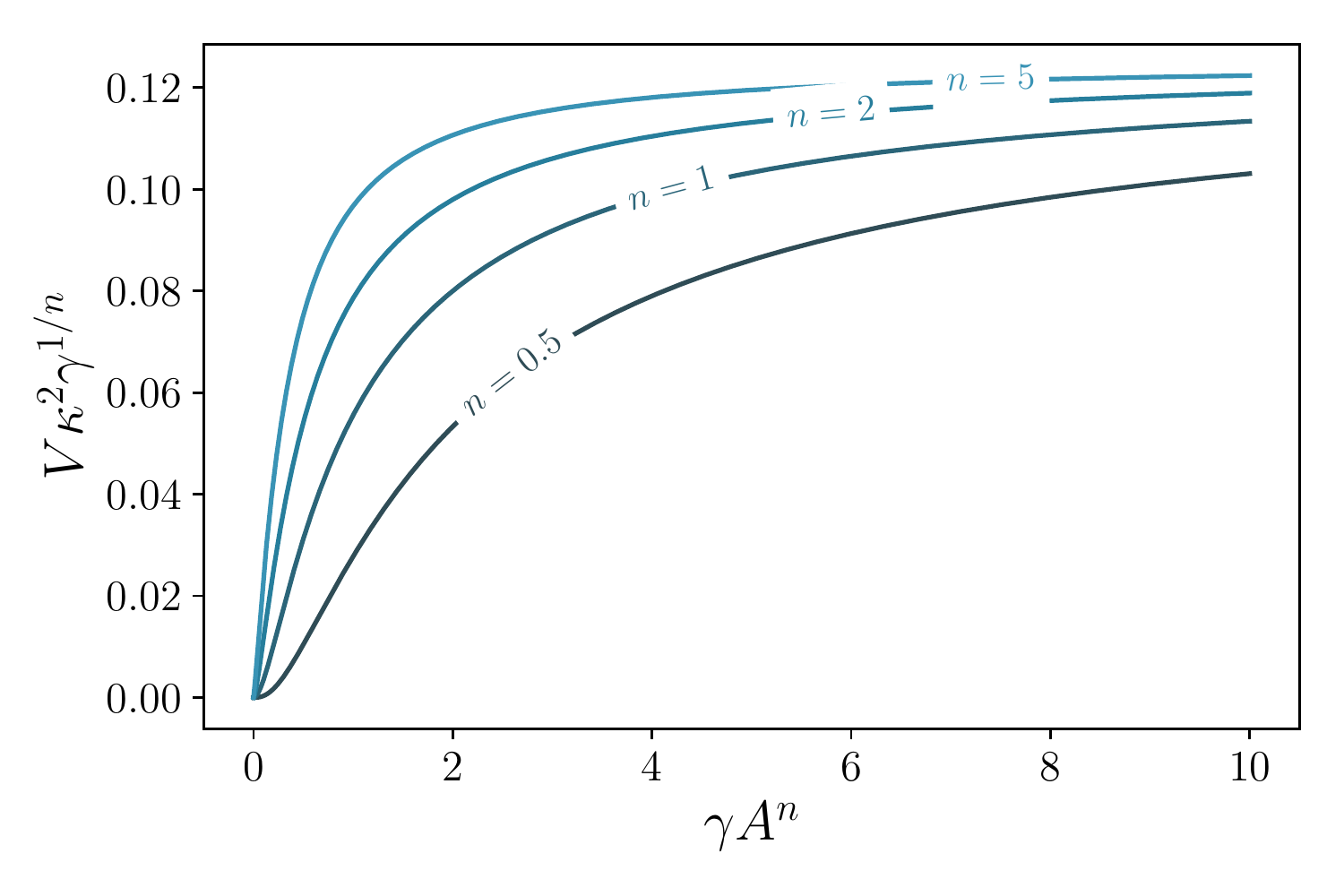}
	\caption{Behavior of the potential for $R(1+\gamma R^n)^{1/n}$ model. }
	\label{images:power_law}
\end{figure}
As is shown in \autoref{images:power_law},
the potential monotonically increases toward a value, $V\kappa^{2}\gamma^{1/n}\sim 1/8$, and a plateau region is present at the large $A$ limit for a positive $n$.
For $-1 \leq n < 0$, however, this model behave as $F = R^2$.
Thus we assume non-zero positive value for $n$.

The integration in \autoref{eq:e-folds-fr} is analytically performed in this model
and the e-folding number is given by
\begin{align}
	\begin{aligned}
		N =& \frac{3\gamma (A_N^n - A_e^n)}{2n} %\\
		+ \frac{3(n-1)}{2n} \ln \abs{\frac{1+\gamma A_N^n}{1+\gamma A_e^n}}
		- \frac{3}{4} \ln \abs{\frac{1+2\gamma A_N^n}{1+2\gamma A_e^n}}.
	\end{aligned}
	\label{e-fold:modified}
\end{align}
Since the first term of the right hand side in \autoref{e-fold:modified}  has a dominant contribution 
at the horizon crossing, we neglect the second and third terms. 
The field variable $A_N$ is much larger than that at the end of inflation, $A_N \gg A_e$. 
Thus  we drop the $A_e$ dependence and find an approximate expression,
\begin{align}
    N \sim 3\gamma A_N^n/2n .
    \label{N:modified}
\end{align} 
Substituting \autoref{N:modified} into Eqs.~\eqref{obs:power_spectrum_fr}--\eqref{obs:non_gaussianity_fr}, we obtain
\begin{align}
	&A_s = \frac{1}{16\pi^2 \gamma^{1/n}}
	\frac{x^{3+1/n} (1 + x)^{1-1/n}}{(1 + 2x)^2}, \\
	&n_s = 1 - \frac{2}{3}
	\frac{(n-1)x^2 + (3n+1)(x+1)^2}{x^2 (1+n+2x)}, \\
	&r = \frac{12}{n^2 N^2}, \\
	&\begin{aligned}[b]
		\alpha_s =& -\frac{8n(2x+1)(x+1)}{9x^4(1+n+2x)^3} \Bigl[ 3n^2(x+1) %\\
				  + n(4x^3+12x^2+13x+4) + (2x+1)^2 \Bigr],
	\end{aligned} \\
	&\begin{aligned}
		f_{\mathrm{NL}} = -\frac{10}{9} \frac{2n(2x-1)(x+2)-6(2x+1)}{x^2 (1+n+2x)},
	\end{aligned}
\end{align}
where $x$ is defined by $x\equiv 2Nn/3$.
These results are simplified to the ones in the Starobinsky inflation Eq.~\eqref{eq:starobinsky-obs} for $n = 1$.

\begin{figure}[htbp]
	\centering
	\includegraphics[width=0.9\linewidth]{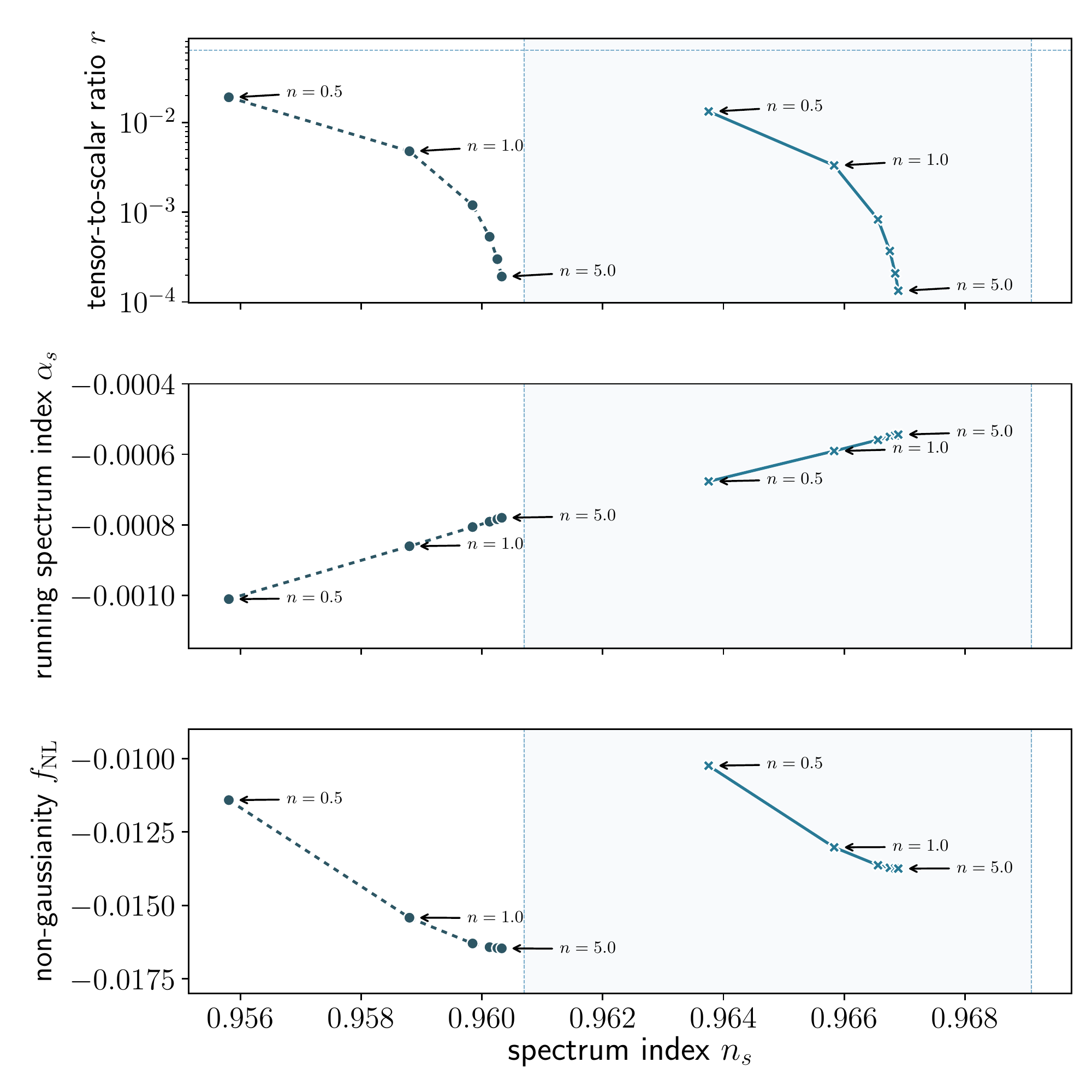}
	\caption{inflationary parameters \autoref{eq:model-toy} from $n=0.5$ to $n=5.0$ for $N=50$ (dashed dark blue lines) and $N=60$ (solid blue lines).}
	\label{images:toy-obs}
\end{figure}

The behavior of the inflationary parameters are depicted in \autoref{images:toy-obs}.
The tensor-to-scalar ratio and the non-gaussianity increase as the parameter $n$ decreases.
As is seen in \autoref{images:toy-obs} the predicted inflationary parameters for $N=60$ satisfy the constraint from the Planck 2018.

\section{Concluding remarks}
\label{sec:concludion}
We have developed useful formulae to calculate the inflationary parameters under the slow-roll scenario in $F(R)$ gravity models. By the Weyl transformation the $F(R)$ gravity model can be described by an equivalent scalar-tensor theory and the scalar field has played a role of the inflaton.

On some classes of the $F(R)$ gravity models are represented by scalar-tensor theories with a non-canonical kinetic term. In our formulae the non-canonical kinetic term is treated by the covariant approach.
The explicit expressions of the inflationary parameters have been obtained as functions of the inflation potential and its derivatives under a general definition of the slow-roll parameters \autoref{eq:potential_slow_roll_1st}, \autoref{eq:potential_slow_roll_2nd}, and \autoref{eq:potential_slow_roll_3rd}.
Thus, the inflationary parameters are derived as a function of $F(R)$ and its derivatives. Evaluating \autoref{eq:e-folds-fr}, we obtain the field variable at the horizon crossing and find the consistent model parameter with current measurements of the inflationary parameters.

We have demonstrated our formulae in the power-law model and the modified power-law model. These models reduce to the Starobinsky model for a specific model parameter. In the power-law model with $N=60$ the consistent parameter is found around the Starobinsky model which can be understood as a model deformed by the quantum corrections. We introduce the modified power-law model as an example with the non-canonical kinetic term. In the modified power-law model the constraints from the inflationary parameters are satisfied for a larger parameter range for $N=60$. The numerical results of the possible parameter space are rendered in \autoref{images:power-law-obs} and \autoref{images:toy-obs}.

Our formulae can be applied to a wide class of $F(R)$ gravity and systematically inspect the consistency with the current observations.
For example, our approach permits to study a general higher derivative correction of gravity with $R^2$ term where the inflation with account of quantum corrections has been studied in Refs.~\refcite{Barrow:1988xh,Barrow:1991hg,Myrzakulov:2014hca}.
It is also interesting to study a model to solve the dark energy, e.g. Refs.~\refcite{Clifton:2005aj,Nojiri:2006ri,Nojiri:2010wj,Clifton:2011jh,Nojiri:2017ncd,Capozziello:2019cav}, and dark matter, e.g. Refs~\refcite{Katsuragawa:2016yir,Katsuragawa:2017wge,Inagaki:2019dqv} problems simultaneously.
The application of covariant approach to $F(R, \phi)$ gravity inflation model, e.g. Refs.~\refcite{Mathew:2017lvh,Johnson:2017pwo,Canko:2019mud} is also worth investigating.
We hope to report some results for these topics in future.

\section*{Acknowledgement}
The authors would like to thank
Y.~Matsuo for valuable discussions.

\bibliography{../refs/refs.bib}

\begin{thebibliography}{10}

\bibitem{Langlois:2010xc}
D.~Langlois, {\em Lect. Notes Phys.} {\bf 800}  (2010) 1,
  \href{http://arxiv.org/abs/1001.5259}{{\ttfamily arXiv:1001.5259
  [astro-ph.CO]}}.

\bibitem{Linde:2014nna}
A.~Linde, { {Inflationary Cosmology after Planck 2013}}, in {\em {Proceedings,
  100th Les Houches Summer School: Post-Planck Cosmology: Les Houches, France,
  July 8 - August 2, 2013}\/},  (2015), pp. 231--316.
\newblock \href{http://arxiv.org/abs/1402.0526}{{\ttfamily arXiv:1402.0526
  [hep-th]}}.

\bibitem{Baumann:2014nda}
D.~Baumann and L.~McAllister, {\em {Inflation and String Theory}}Cambridge
  Monographs on Mathematical Physics, Cambridge Monographs on Mathematical
  Physics (Cambridge University Press, 2015).

\bibitem{Clifton:2005aj}
T.~Clifton and J.~D. Barrow, {\em Phys. Rev.} {\bf D72}  (2005)   103005,
  \href{http://arxiv.org/abs/gr-qc/0509059}{{\ttfamily arXiv:gr-qc/0509059
  [gr-qc]}}, [Erratum: Phys. Rev.D90,no.2,029902(2014)].

\bibitem{Nojiri:2006ri}
S.~Nojiri and S.~D. Odintsov, {\em eConf} {\bf C0602061}  (2006)  ~06,
  \href{http://arxiv.org/abs/hep-th/0601213}{{\ttfamily arXiv:hep-th/0601213
  [hep-th]}}, [Int. J. Geom. Meth. Mod. Phys.4,115(2007)].

\bibitem{Nojiri:2010wj}
S.~Nojiri and S.~D. Odintsov, {\em Phys. Rept.} {\bf 505}  (2011) 59,
  \href{http://arxiv.org/abs/1011.0544}{{\ttfamily arXiv:1011.0544 [gr-qc]}}.

\bibitem{Clifton:2011jh}
T.~Clifton, P.~G. Ferreira, A.~Padilla and C.~Skordis, {\em Phys. Rept.} {\bf
  513}  (2012) 1, \href{http://arxiv.org/abs/1106.2476}{{\ttfamily
  arXiv:1106.2476 [astro-ph.CO]}}.

\bibitem{Nojiri:2017ncd}
S.~Nojiri, S.~D. Odintsov and V.~K. Oikonomou, {\em Phys. Rept.} {\bf 692}
  (2017) 1, \href{http://arxiv.org/abs/1705.11098}{{\ttfamily arXiv:1705.11098
  [gr-qc]}}.

\bibitem{Capozziello:2019cav}
S.~Capozziello, R.~D'Agostino and O.~Luongo, {\em Int. J. Mod. Phys.} {\bf D28}
   (2019)   1930016, \href{http://arxiv.org/abs/1904.01427}{{\ttfamily
  arXiv:1904.01427 [gr-qc]}}.

\bibitem{Starobinsky:1980te}
A.~A. Starobinsky, {\em Phys. Lett.} {\bf 91B}  (1980) 99, [,771(1980)].

\bibitem{Akrami:2018odb}
 Planck Collaboration (Y.~Akrami {\em et~al.})  (2018)
  \href{http://arxiv.org/abs/1807.06211}{{\ttfamily arXiv:1807.06211
  [astro-ph.CO]}}.

\bibitem{Akrami:2019izv}
 Planck Collaboration (Y.~Akrami {\em et~al.})  (2019)
  \href{http://arxiv.org/abs/1905.05697}{{\ttfamily arXiv:1905.05697
  [astro-ph.CO]}}.

\bibitem{Sasaki:1995aw}
M.~Sasaki and E.~D. Stewart, {\em Prog. Theor. Phys.} {\bf 95}  (1996) 71,
  \href{http://arxiv.org/abs/astro-ph/9507001}{{\ttfamily
  arXiv:astro-ph/9507001 [astro-ph]}}.

\bibitem{Peterson:2011yt}
C.~M. Peterson and M.~Tegmark, {\em Phys. Rev.} {\bf D87}  (2013)   103507,
  \href{http://arxiv.org/abs/1111.0927}{{\ttfamily arXiv:1111.0927
  [astro-ph.CO]}}.

\bibitem{Gao:2014fva}
X.~Gao, T.~Li and P.~Shukla, {\em JCAP} {\bf 1410}  (2014)   008,
  \href{http://arxiv.org/abs/1403.0654}{{\ttfamily arXiv:1403.0654 [hep-th]}}.

\bibitem{Elliston:2012ab}
J.~Elliston, D.~Seery and R.~Tavakol, {\em JCAP} {\bf 1211}  (2012)   060,
  \href{http://arxiv.org/abs/1208.6011}{{\ttfamily arXiv:1208.6011
  [astro-ph.CO]}}.

\bibitem{Brooker:2016oqa}
D.~J. Brooker, S.~D. Odintsov and R.~P. Woodard, {\em Nucl. Phys.} {\bf B911}
  (2016) 318, \href{http://arxiv.org/abs/1606.05879}{{\ttfamily
  arXiv:1606.05879 [gr-qc]}}.

\bibitem{Schwarz:2001vv}
D.~J. Schwarz, C.~A. Terrero-Escalante and A.~A. Garcia, {\em Phys. Lett.} {\bf
  B517}  (2001) 243, \href{http://arxiv.org/abs/astro-ph/0106020}{{\ttfamily
  arXiv:astro-ph/0106020 [astro-ph]}}.

\bibitem{Gong:2011uw}
J.-O. Gong and T.~Tanaka, {\em JCAP} {\bf 1103}  (2011)   015,
  \href{http://arxiv.org/abs/1101.4809}{{\ttfamily arXiv:1101.4809
  [astro-ph.CO]}}, [Erratum: JCAP1202,E01(2012)].

\bibitem{Gong:2014kpa}
J.-O. Gong, {\em JCAP} {\bf 1505}  (2015)   041,
  \href{http://arxiv.org/abs/1409.8151}{{\ttfamily arXiv:1409.8151
  [astro-ph.CO]}}.

\bibitem{Sebastiani:2013eqa}
L.~Sebastiani, G.~Cognola, R.~Myrzakulov, S.~D. Odintsov and S.~Zerbini, {\em
  Phys. Rev.} {\bf D89}  (2014)   023518,
  \href{http://arxiv.org/abs/1311.0744}{{\ttfamily arXiv:1311.0744 [gr-qc]}}.

\bibitem{Motohashi:2014tra}
H.~Motohashi, {\em Phys. Rev.} {\bf D91}  (2015)   064016,
  \href{http://arxiv.org/abs/1411.2972}{{\ttfamily arXiv:1411.2972
  [astro-ph.CO]}}.

\bibitem{Liu:2018htf}
L.-H. Liu  (2018) \href{http://arxiv.org/abs/1807.00666}{{\ttfamily
  arXiv:1807.00666 [gr-qc]}}.

\bibitem{Ben-Dayan:2014isa}
I.~Ben-Dayan, S.~Jing, M.~Torabian, A.~Westphal and L.~Zarate, {\em JCAP} {\bf
  1409}  (2014)   005, \href{http://arxiv.org/abs/1404.7349}{{\ttfamily
  arXiv:1404.7349 [hep-th]}}.

\bibitem{Odintsov:2017hbk}
S.~D. Odintsov, V.~K. Oikonomou and L.~Sebastiani, {\em Nucl. Phys.} {\bf B923}
   (2017) 608, \href{http://arxiv.org/abs/1708.08346}{{\ttfamily
  arXiv:1708.08346 [gr-qc]}}.

\bibitem{Barrow:1988xh}
J.~D. Barrow and S.~Cotsakis, {\em Phys. Lett.} {\bf B214}  (1988) 515.

\bibitem{Barrow:1991hg}
J.~D. Barrow and S.~Cotsakis, {\em Phys. Lett.} {\bf B258}  (1991) 299.

\bibitem{Myrzakulov:2014hca}
R.~Myrzakulov, S.~Odintsov and L.~Sebastiani, {\em Phys. Rev.} {\bf D91}
  (2015)   083529, \href{http://arxiv.org/abs/1412.1073}{{\ttfamily
  arXiv:1412.1073 [gr-qc]}}.

\bibitem{Katsuragawa:2016yir}
T.~Katsuragawa and S.~Matsuzaki, {\em Phys. Rev.} {\bf D95}  (2017)   044040,
  \href{http://arxiv.org/abs/1610.01016}{{\ttfamily arXiv:1610.01016 [gr-qc]}}.

\bibitem{Katsuragawa:2017wge}
T.~Katsuragawa and S.~Matsuzaki, {\em Phys. Rev.} {\bf D97}  (2018)   064037,
  \href{http://arxiv.org/abs/1708.08702}{{\ttfamily arXiv:1708.08702 [gr-qc]}},
  [Erratum: Phys. Rev.D97,no.12,129902(2018)].

\bibitem{Inagaki:2019dqv}
T.~Inagaki, Y.~Matsuo and H.~Sakamoto  (2019)
  \href{http://arxiv.org/abs/1905.05503}{{\ttfamily arXiv:1905.05503 [gr-qc]}}.

\bibitem{Mathew:2017lvh}
J.~Mathew, J.~P. Johnson and S.~Shankaranarayanan, {\em Gen. Rel. Grav.} {\bf
  50}  (2018)  ~90, \href{http://arxiv.org/abs/1705.07945}{{\ttfamily
  arXiv:1705.07945 [gr-qc]}}.

\bibitem{Johnson:2017pwo}
J.~P. Johnson, J.~Mathew and S.~Shankaranarayanan, {\em Gen. Rel. Grav.} {\bf
  51}  (2019)  ~45, \href{http://arxiv.org/abs/1706.10150}{{\ttfamily
  arXiv:1706.10150 [gr-qc]}}.

\bibitem{Canko:2019mud}
D.~D. Canko, I.~D. Gialamas and G.~P. Kodaxis  (2019)
  \href{http://arxiv.org/abs/1901.06296}{{\ttfamily arXiv:1901.06296
  [hep-th]}}.

\end{thebibliography}
\end{document}